\documentclass[a4paper]{jpconf}
\usepackage{graphicx}
\usepackage[small]{subfigure}

\begin{document}
\title{Commissioning of the CALIFA Barrel Calorimeter of the R$^{3}$B Experiment at FAIR}

\author{P Cabanelas$^{1,2}$, H Alvarez-Pol$^{1,2}$, J M Boillos$^{1,2}$, E Casarejos$^{3}$, J Cederkall$^{4}$, D Cortina$^{1,2}$, M Feijoo$^{1,2}$, D Galaviz$^{5,6}$, E Galiana$^{1,5}$, R Gernh\"auser$^{7}$, P Golubev$^{4}$, D Gonz\'alez$^{1,2}$, A-L Hartig$^{8}$, A Heinz$^{9}$, H Johansson$^{9}$ P Klenze$^{7}$, A Knyazev$^{4}$, T Kr\"oll$^{8}$, E Nacher$^{10}$, J Park$^{4}$, A Perea$^{11}$, L Ponnath$^{7}$, H-B Rhee$^{8}$, J L Rodr\'iguez-S\'anchez$^{1,2}$, C Suerder$^{8}$, O Tengblad$^{11}$ and P Teubig$^{5}$  for the CALIFA Working Group of the R$^3$B Experiment}

\address{$^{1}$ Instituto Galego de F\'isica de Altas Enerx\'ias (IGFAE), E-15782 Universidade de Santiago de Compostela, Spain}
\address{$^{2}$ Particle Physics Department, University of Santiago de Compostela, E-15782 Santiago de Compostela, Spain}
\address{$^{3}$ Universidade de Vigo, E-36310 Vigo, Spain}
\address{$^{4}$ Department of Physics, Lund University, SE-221 00 Lund, Sweden}
\address{$^{5}$ Laboratory for Instrumentation and Experimental Particle Physics, LIP, 1649-003 Lisbon, Portugal}
\address{$^{6}$ Departamento de F\'isica, Faculdade de Ci\^encias da Universidade de Lisboa, 1749-016 Lisbon, Portugal}
\address{$^{7}$ Physik Department, Technische Universit\"at M\"unchen, 85748 Garching, Germany}
\address{$^{8}$ Institut f\"ur Kernphysik, Technische Universit\"at Darmstadt, D-64289 Darmstadt, Germany}
\address{$^{9}$ Institutionen f\"or Fysik, Chalmers Tekniska H\"ogskola, S-412 96 G\"oteborg, Sweden}
\address{$^{10}$ Instituto de F\'isica Corpuscular, Universitat de Valencia, E-46980, Valencia, Spain}
\address{$^{11}$ Instituto de Estructura de la Materia, CSIC, E-28006 Madrid, Spain}

\ead{pablo.cabanelas@usc.es}

\begin{abstract}
CALIFA is the high efficiency and energy resolution calorimeter for the R$^{3}$B experiment at FAIR, intended for detecting high energy charged particles and $\gamma$-rays in inverse kinematics direct reactions. It surrounds the reaction target in a segmented configuration of Barrel and Forward End-Cap pieces. The CALIFA Barrel consists of 1952 detection units made of CsI(Tl) long-shaped scintillator crystals, and it is being commissioned during the Phase0 experiments at FAIR. The first setup for the CALIFA Barrel commissioning is presented here. Results of detector performance with $\gamma$-rays are obtained, and show that the system fulfills the design requirements.
\end{abstract}

\section{Introduction}
CALIFA (CALorimeter for In-Flight detection of gamma-rays and high energy charged pArticles) \cite{Cortina,califaSim}, is the calorimeter detector of the R$^{3}$B (Reactions with Relativistic Radioactive Beams) experiment at FAIR (Facility for Anti-proton and Ion Research), Darmstadt, Germany. It consists of 2432 detection units of long CsI Tl-doped scintillator crystals, with Large Area Avalanche Photo-Diode (LAAPD, or simply APD) based readout, arranged in a barrel (1952 detection units) plus forward end-cap (480 detection units) structure surrounding the target area, covering a polar angle from 19$^\mathrm{o}$ to 140$^\mathrm{o}$, full azimuthal coverage, a huge detection dynamic range from around 100 keV gamma rays to 320 MeV protons and a foreseen energy resolution below 6\% at 1 MeV after event reconstruction. The lowest polar angle region of End-Cap is covered by a LaBr$_3$/LaCl$_3$ phoswich array section called CEPA.

The Barrel section is already under construction and some fractions were recently commissioned during the first Phase0 experiments at FAIR/GSI. A brief description of the CALIFA Barrel system and the description and first results of the commissioning are presented in this document.

\section{The CALIFA Barrel System}
\begin{figure}[htb!]
\centering
\subfigure[]{\includegraphics[width=0.40\textwidth]{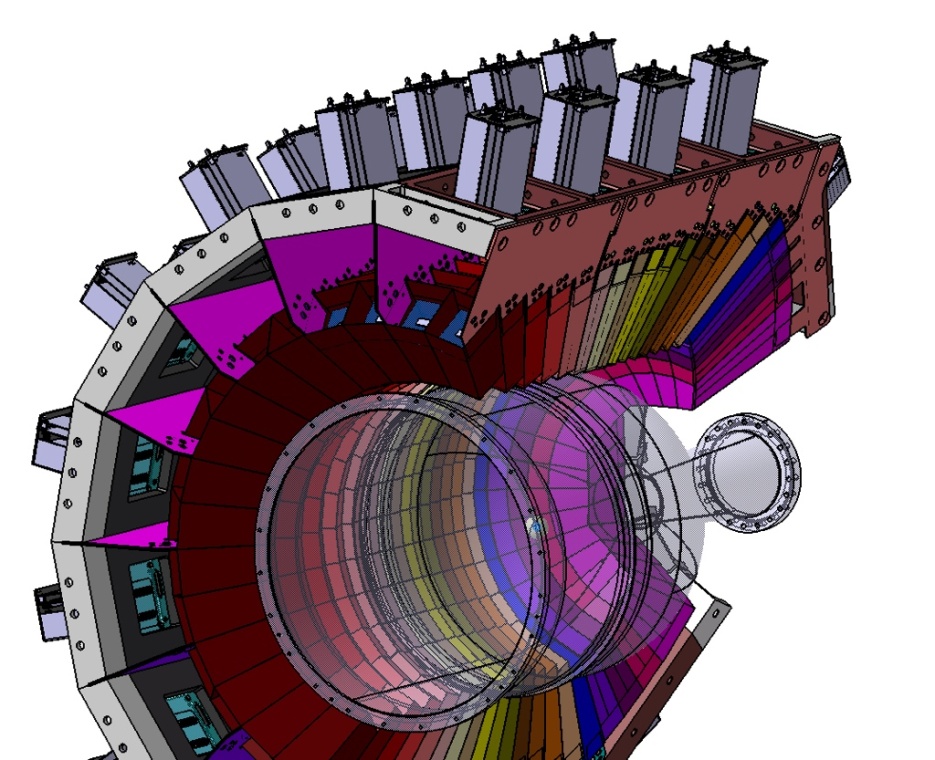}}
\subfigure[]{\includegraphics[width=0.55\textwidth]{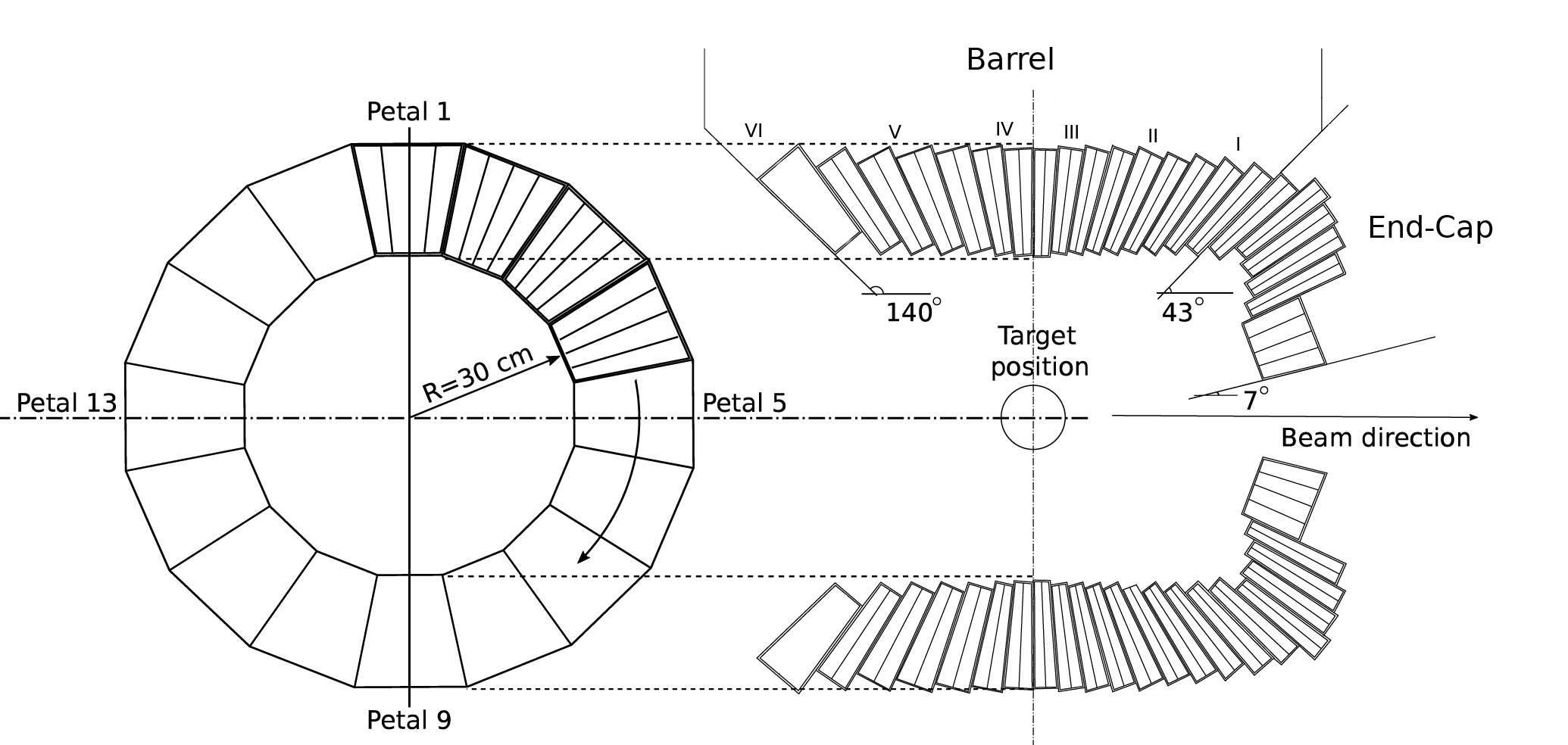}}
\caption{(a) Artistic view of the CALIFA detector system together with the reaction Chamber of R3B. The carbon fiber alveoli and aluminum holders for the crystals are drawn, as well as the preamplifiers (grey boxes) surrounding the structure. (b) Layout of CALIFA detector. These figures do not represent crystals, but rather alveoli, although it is clear how all crystals inside alveoli are pointing to the interaction area. Each alveolus contains 4 individual wrapped crystals, except for the furthermost backward angle ring, containing a single crystal. Figure (b) adapted from \cite{lund}.} \label{3D_and_layout_fig}
\end{figure}

The so-called Barrel is the CALIFA section covering polar angles between 43$^\mathrm{o}$ to 140$^\mathrm{o}$, with full azimuthal coverage, which corresponds roughly to the 40\% of $\gamma$-rays emitted in a typical R3B reaction. The Barrel, a cylindrical pattern around the beam line, has an inner radius of 30 cm, and comprises a total of 1952 detection units (CsI(Tl) crystals + LAAPD) of 11 different geometries arranged in carbon fiber support assemblies called alveoli. These geometries can be classified in six main groups: groups I, II and III are made of 3 rings of detection units, and the crystal lengths are, respectively, 220 mm, 180 mm and 170 mm; group IV comprises 2 rings, and crystals are 160 mm long; group V is constituted by 4 rings and crystal lengths are 140 mm; finally, group VI corresponds to the furthermost backward ring, where each alveolus contains a single crystal of 120 mm. The Barrel design, granularity and crystals location has been determined for a constant relative resolution of around 6\% for $\gamma$-rays at 1 MeV, high detection efficiency and full geometrical coverage, minimizing losses due to partial detection of Compton events, and allows for the detection of protons, i.e., in the case of \emph{(p,2p)} reactions. The longitudinal axis of all crystals points to the interaction area at any time. Figure \ref{3D_and_layout_fig} shows and artistic view of the CALIFA detector and a layout where the region covered by the Barrel, and the different crystals groups are drawn.

\section{The CALIFA Barrel Commissioning} 
The first experiment for the CALIFA Barrel Commissioning within the FAIR Phase0 was scheduled in February 2019 at GSI. Seven segments, or petals, of the CALIFA Barrel were completely installed and instrumented in a dedicated frame, making a total of 448 detection units. Five petals were installed in their nominal positions, while the other two were slightly tilted in the forward direction in order to increase the geometry detection efficiency for the \emph{(p,2p)} reactions. Each petal enclosed 64 CsI(Tl) scintillator crystals, of three different lengths, 220 mm, 180 mm and 170 mm, together with their corresponding APDs. The petals were located around the interaction point covering different regions of azimuthal angle, and covering polar angle regions between 85$^\mathrm{o}$ and 35$^\mathrm{o}$ approximately. They were instrumented with their final Mesytec MPRB-32 pre-amplifiers, which were operated some in \emph{single} mode and some in \emph{dual} mode, that is, two different read-outs with different gain range for a single data channel. That configuration allowed for the optimal detection of both $\gamma$-rays (high gain) and protons (low gain) in the same crystal. The data acquisition (DAQ) and trigger logic was made with the FEBEX-based DAQ and trigger system \cite{Febex}, which will be also the final DAQ system for CALIFA, and the event synchronization with the other R3B detection systems was carried out by means of the White Rabbit timing distribution protocol network \cite{whiterabbit}. In addition, an online monitoring system was developed for the CALIFA detector based on the online features of the R3BRoot framework \cite{r3broot}, and was successfully used during the commissioning for the first time.

\begin{figure}[htb!]
\centering
\subfigure[]{\includegraphics[width=0.45\textwidth]{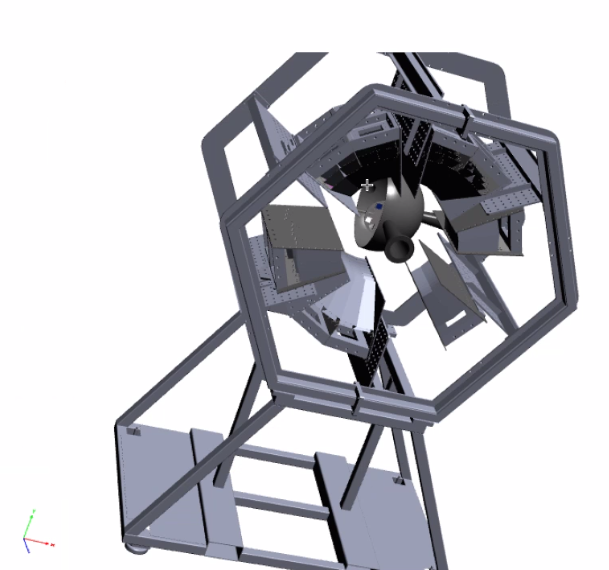}}
\subfigure[]{\includegraphics[width=0.45\textwidth]{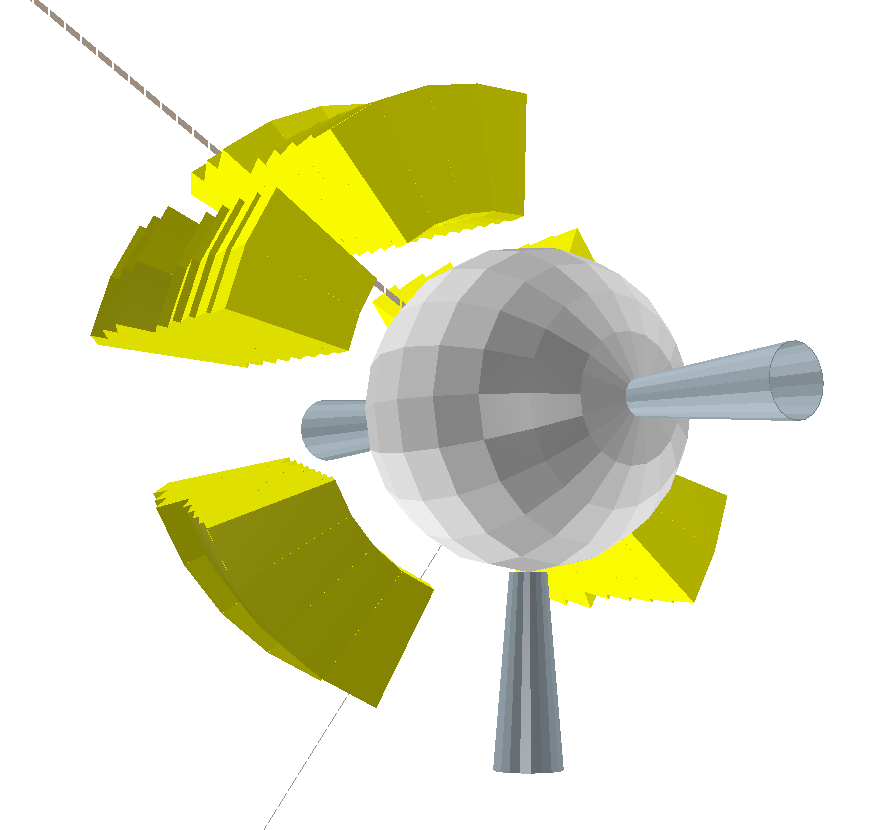}}
\caption{(a) Complete drawing of the mechanical structure for the commissioning with the dedicated frame, the seven petals surrounding the interaction area and the spherical reaction chamber. (b) View of simulated event over the detection units arranged in the commissioning configuration.} \label{sim_setup_fig}
\end{figure}

\begin{figure}[htb!]
\centering
\includegraphics[width=0.8\textwidth]{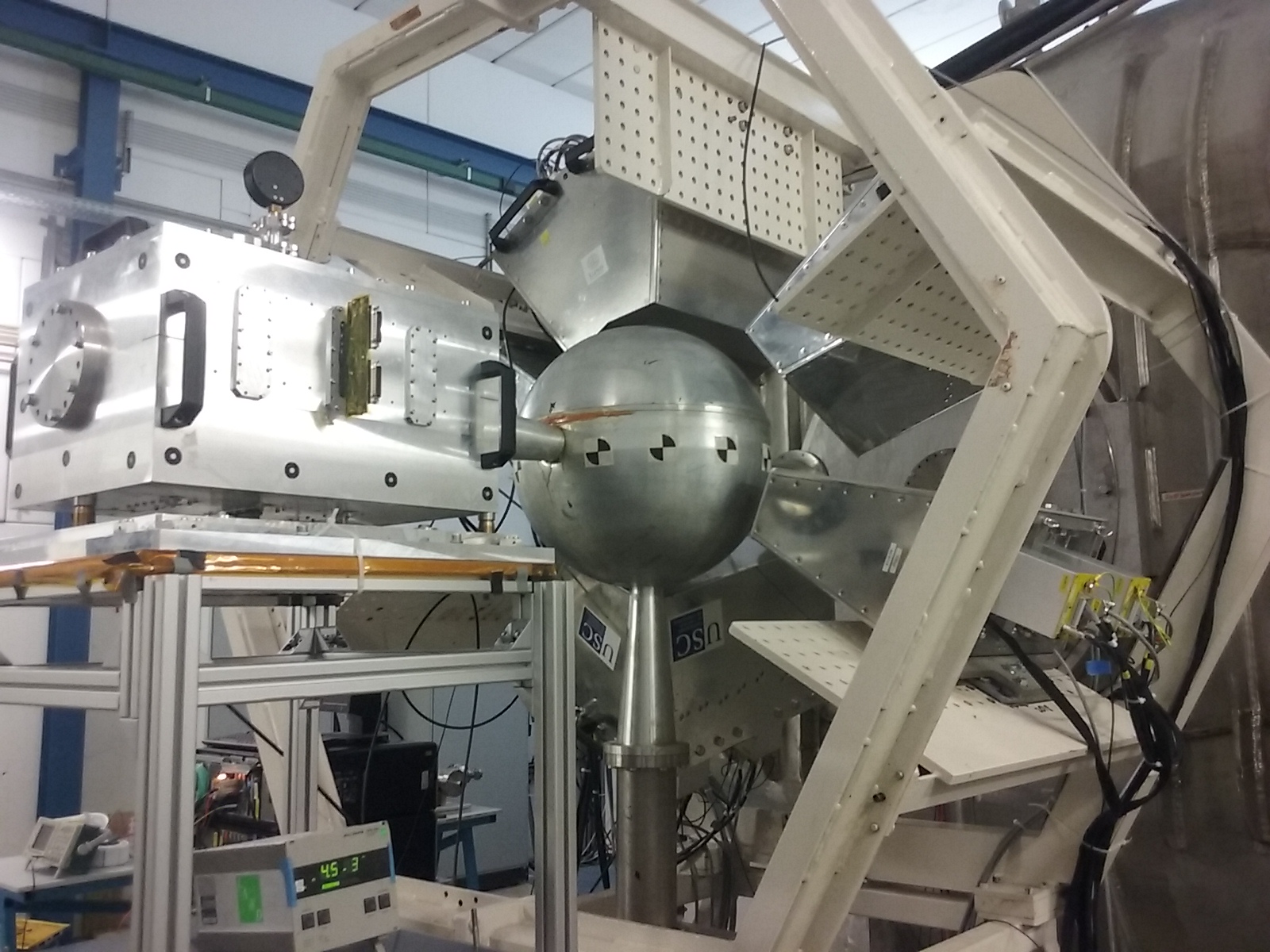}
\caption{Photograph of the CALIFA Barrel commissioning setup: the 7 Barrel petals are attached to the dedicated frame (white steel structure), located in front of the GLAD R3B magnet. The spherical reaction chamber in the interaction point is surrounded by the petals, and a multi-wire tracking detector was placed in the beam-line before the interaction point.} \label{real_setup_fig}
\end{figure}

Figure \ref{sim_setup_fig} shows a drawing of the mechanical structure for the commissioning, together with the petals and reaction chamber (a), and a view of a simulated event (b) in the implementation done in R3BRoot for a further analysis with simulated data. Figure \ref{real_setup_fig} shows the photograph of the final CALIFA Barrel commissioning setup. The 7 petals are fully instrumented and attached to the frame, in front of the R3B-GLAD superconducting magnet. A spherical reaction chamber was placed as well in the interaction area, housing the target wheel and silicon tracking detectors. Before the target point, upstream in the beam line, a multi-wire detector was also placed. 

We measured during 7 days of non-continuous proton and $^{12}$C beam from super-FRS accelerator, over a CH$_{2}$ segmented target. Interesting Physics of the quasi-elastic \emph{(p,2p)} reaction produced is being currently analyzed. It is well worthy to mention that all installed detection units were full operative during the whole beam-time, that is, 448 data channels were continuously read-out. 

Calibrations runs were taken prior to the beam-time with a $^{60}$Co radioactive source placed in the interaction point. Thus, all detection units were illuminated with $\gamma$-rays at the same time. Data was collected with the system mentioned above, and the performances were measured with those $\gamma$-rays. The calibration runs also allowed for the so-called \emph{gain matching} procedure \cite{lund}, a fine tuning of the preamplifiers HV bias so that all channels show the photopeaks in the same spectrum region.  

Once a given spectrum is recorded, the corresponding photopeaks were automatically searched and analyzed with the TSpectrum feature of ROOT \cite{ROOT}, and, after a background fitting and subtraction procedure, a two step recursive and automatic Gaussian fit was applied in order to obtain the mean and the sigma $\sigma$ of the photopeak. The energy resolution is calculated by $Res[\%] = \frac{\Delta E}{E} \times 100 = \frac{2.35\sigma}{E} \times 100$, being $2.35\sigma$ the full width at half maximum (FWHM). The final value of energy resolution at 1 MeV is calculated by fitting the corresponding energy resolution values of measured photopeaks to a function $Res(E) = \frac{a}{\sqrt{E}} + b$ \cite{knoll}.

\begin{figure}[htb!]
\centering
\subfigure[]{\includegraphics[width=0.49\textwidth]{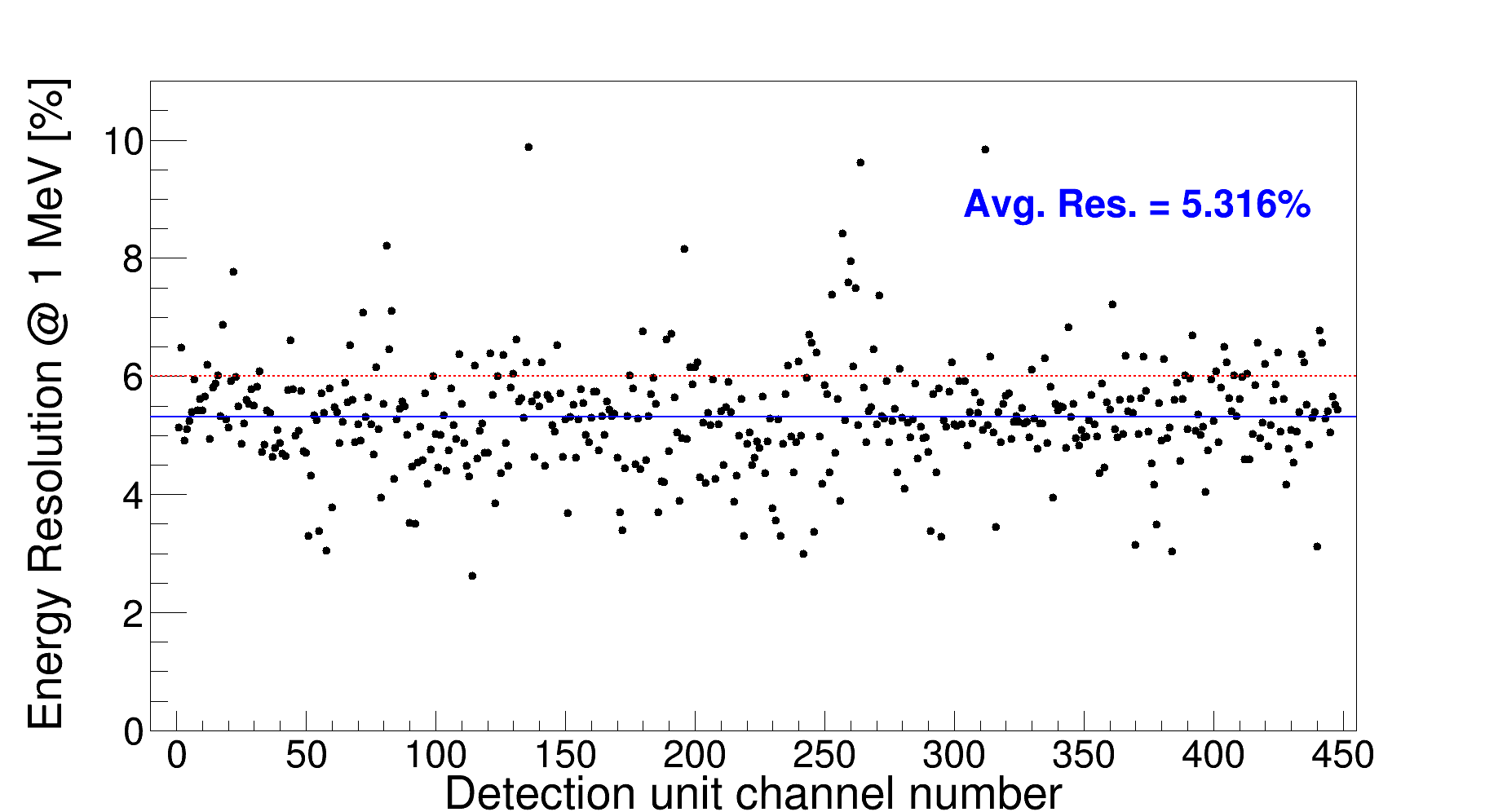}}
\subfigure[]{\includegraphics[width=0.49\textwidth]{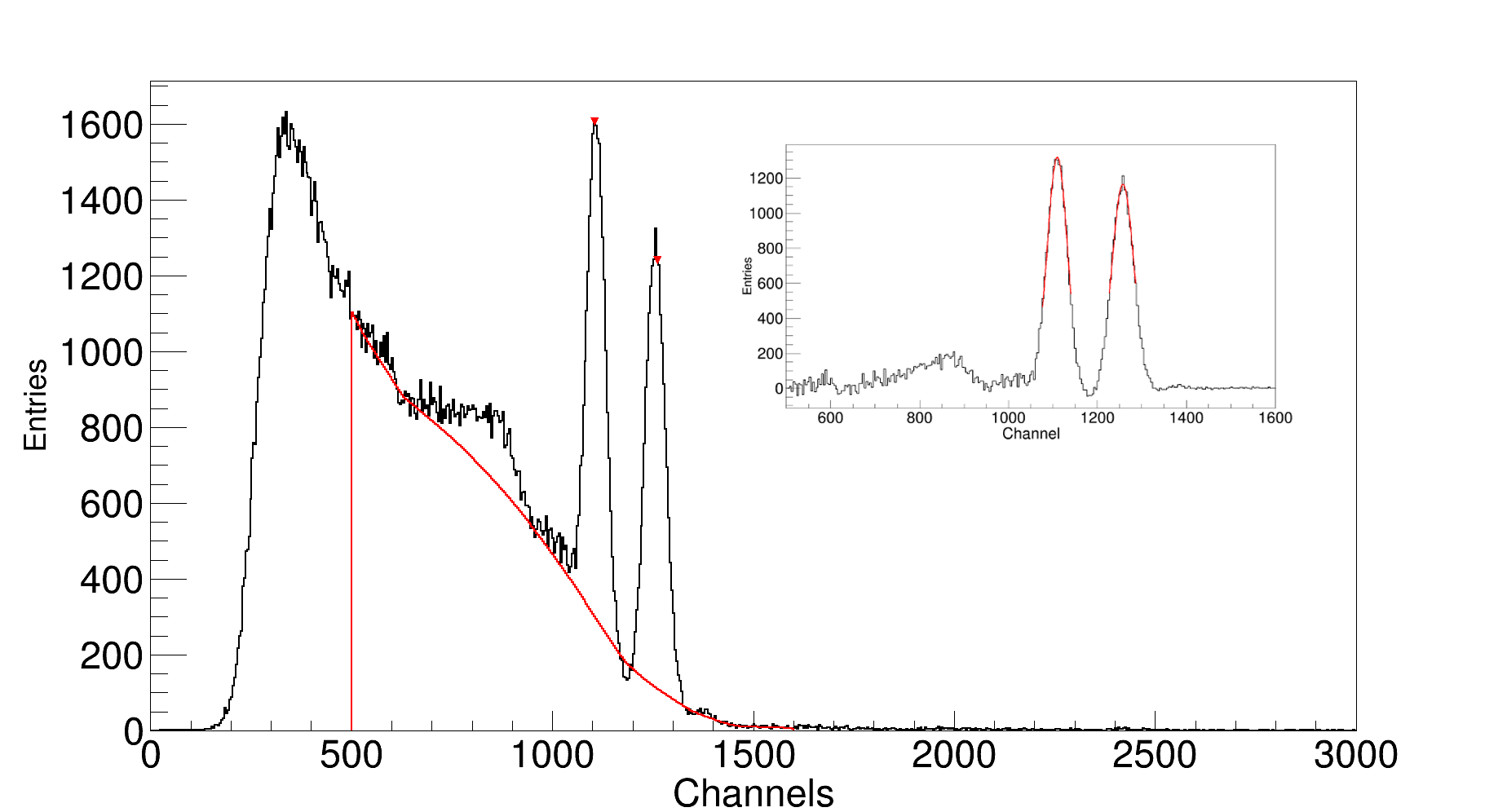}}
\caption{ (a) Energy resolution at 1 MeV for the 448 instrumented channels. The red dashed line corresponds to the required 6\% resolution, while continuous blue line shows the obtained average energy resolution of 5.316\%. (b) Example of a typical recorded spectrum showing also the background line (in red), and the fit to both $^{60}$Co photopeaks after background subtraction (top right corner insert).} \label{results_fig}
\end{figure}

Figure \ref{results_fig} (a) shows all the energy resolution values at 1 MeV measured for the commissioning setup with a $^{60}$Co source placed in the interaction point. The average energy resolution was obtained to be 5.316\%, fulfilling the system requirements of 6\%, and most of the channels showed a value around 5\%. No significant differences were observed in terms of resolution for the different crystal geometries and sizes. Figure \ref{results_fig} (b) represents a typical recorded calibration spectrum with the estimated background and the Gaussian fit to the $^{60}$Co 1.17 and 1.33 MeV photopeaks after background subtraction.

\section{Conclusions}
In February 2019, seven segments, or petals, of the CALIFA Barrel calorimeter for R3B have been commissioned for the first time at GSI/FAIR. A total of 448 detection units were installed in a dedicated structure and instrumented with the final Front-End Electronics and Data Acquisition systems. Data was taken with radioactive sources and also under beam conditions. All detection units were successfully operated during the experiment. An average resolution for the system of 5.316\% for $\gamma$-rays at 1 MeV was obtained. Analysis \emph{(p,2p)} of beam data with the CALIFA petals is ongoing. 

\ack The author wish to acknowledge encouragement from all the team involved in the CALIFA Barrel commissioning experiments development. The author's work has been financially supported by the Spanish MICCIN grant FPA2015-69640-C2-1-P.

\section*{References}

\bibliography{mybibfile}

\end{document}